\documentclass[12pt]{article}

\usepackage{graphicx}
\usepackage{booktabs} 
\usepackage{amsmath}
\begin{document}

\title{An LSTM-Based Dynamic Customer Model for Fashion Recommendation}

\author{Sebastian Heinz, Christian Bracher, Roland Vollgraf\\ Zalando Research\\ \small{\{christian.bracher, sebastian.heinz, roland.vollgraf\}@zalando.de}}

\maketitle

\begin{abstract}
Online fashion sales present a challenging use case for personalized recommendation:  Stores offer a huge variety of items in multiple sizes. Small stocks, high return rates, seasonality, and changing trends cause continuous turnover of articles for sale on all time scales.  Customers tend to shop rarely, but often buy multiple items at once.
We report on backtest experiments with sales data of 100k frequent shoppers at Zalando, Europe's leading online fashion platform. To model changing customer and store environments, our recommendation method employs a pair of neural networks: To overcome the cold start problem, a feedforward network generates article embeddings in ``fashion space,'' which serve as input to a recurrent neural network that predicts a style vector in this space for each client, based on their past purchase sequence.  We compare our results with a static collaborative filtering approach, and a popularity ranking baseline.
\end{abstract}

\section{Introduction}
The recommendation task in the setting of online fashion sales presents unique challenges.  Consumer tastes and body shapes are idiosyncratic, so a huge selection of items in different sizes must be kept on offer.  On a typical day, Zalando, Europe's leading online fashion platform with $\sim$20M active customers, offers $\sim$200k product choices for sale.  Being physical goods rather than digital information, fashion articles must be stocked in warehouses; as most of them are rarely ordered, items are generally available in small, fluctuating numbers. In addition, shoppers commonly return articles.  The result is a rapid turnover of the inventory, with many items going in and out of stock daily.  Superimposed on short-scale variations, there are periodic alterations associated with the seasonal cycle, and secular changes caused by fashion trends.  Regarding consumer behavior, a noteworthy difference to e.g.\ streaming media services is their propensity to buy rarely (a few sales annually), but then multiple items at once.  Hence, their purchase histories are sparse, only partially ordered sequences.

We previously introduced a recommendation algorithm for fashion items that combines article images, tags, and other catalog information with customer response, tethering curated content to collaborative filtering by minimizing the cross-entropy loss of a deep neural network for the sales record across a large selection of customers \cite{kdd}.  Like logistic matrix factorization methods \cite{kdd:matrixfactorization,kdd:logisticfactorization}, our technique yields low-dimensional embeddings for articles (``Fashion DNA'') and customers (``style vectors''), but has the advantage to circumvent the cold-start problem that plagues collaborative methods by injecting catalog information for newly added articles.  Our model proves capable of recognizing individual style preferences from a modest number of purchases; as cumulative sales events extend over a multi-year period, however, it creates only a static style ``fingerprint'' of a customer.

In this contribution, we start from the static model, but extend it by including time-of-sale information.  To contend with the ever-varying article stock, we use the static model to generate Fashion DNA from curated article data, and employ it as a fixed item descriptor.  This allows us to focus on the temporal sequence of sales events for individual customers, which we feed into a neural network to estimate their style vectors.  As these are updated with every purchase, the approach models the evolution of our customers' tastes, and we may employ the style vectors at a given date to create a personalized preference ranking of the articles then in store, in a way fully analogous to the static model.  Recurrent neural networks (RNN) are specifically designed to handle sequential data (see Chapter~10 in Ref.~\cite{deeplearning} for an overview).  Our network, introduced in Section~\ref{sec:dynamicmodel}, employs long short-term memory (LSTM) cells \cite{lstm} to learn temporal correlations between sales.  As the model shares network weights between customers, it has comparatively few parameters, and easily scales to millions of clients during inference.

Recently, evaluations have appeared in the literature \cite{Devooght,Ko, Wang} that indicate superiority of RNN-based recommender systems on standard data sets (LastFM, Netflix) over static models.
Comparing the dynamic customer style model with predictions from the static counterpart \cite{kdd}, and a baseline model build on global customer preferences, we confirm that 
fashion recommendation benefits from temporal information (Section~\ref{sec:comparison}).  However, we also find that peculiarities innate to the fashion context, like the prevalence of partially ordered purchase sequences and the variability of in-store content, are prone to impact recommendation quality; care must be taken in designing RNN architecture, training, and evaluation schemes to accommodate them.  Further avenues for research are discussed in Section~\ref{sec:outlook}.

\section{A dynamic recommender system}
\label{sec:dynamicmodel}
We now lay out the elements of our proposed model -- the data used for training and validation, the static network learning the article embeddings (Fashion DNA), the recurrent network responsible for predicting the customer response, and the training scheme.

\subsection{Data overview}
\label{subsec:data}

This study is based on article and sales data from Zalando's online fashion store, collected from its start in 2008, up to a cutoff date of July 1, 2015.  The data set contains information about $\sim$1M fashion items and millions of individual sales events (excluding customer returns).  Merchandise is characterized by a thumbnail image of each item (size 108$\times$156), categorical data (brand, color, gender, etc.) that has been rolled out into $\sim$7k one-hot encoded ``tags,'' and as numerical data, the logarithm of the manufacturer-suggested retail price, and, for garments only, the fabric composition across $\sim$50 fibers as percentages.  Each sales record contains a unique, anonymized customer ID, the article bought (disregarding size information), and the time of sale, with one minute granularity.  Customer data is limited to sales; in particular, article ratings were not available.

\subsection{Fashion DNA}
\label{subsec:fdna}

Our first task is to encode the properties of the articles in a dense numerical representation.  As the curated data has multiple formats and carries diverse information, a natural vehicle for this transformation is a deep neural network that learns suitable combinations of features on its own.  We discussed such a model at length in an earlier paper \cite{kdd}, and we will only give an overview here.

\begin{figure}
\centering
\includegraphics[width=0.97\columnwidth]{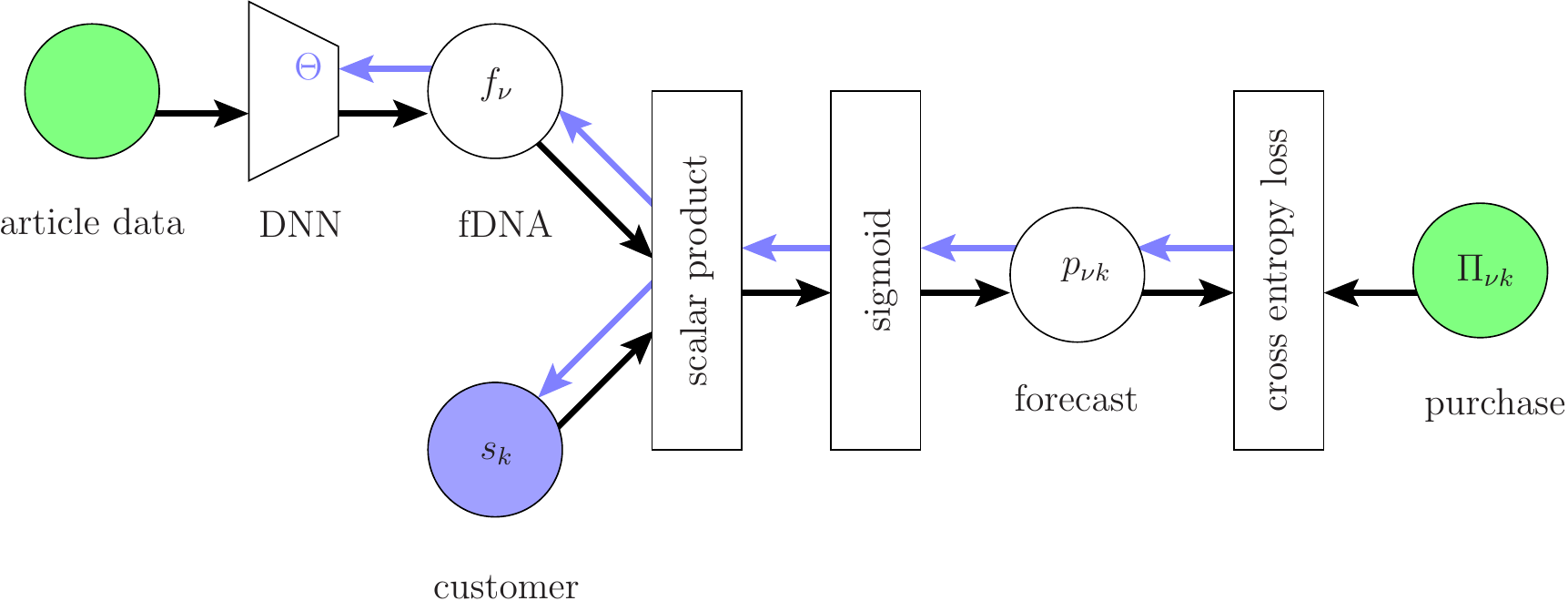}
\caption{Training the Fashion~DNA network. Backpropagation of the loss (blue arrows) simultaneously improves the static customer style vectors $s_k$, and the network weights $\Theta$.}
\label{fig:static_net}
\end{figure}

The representation of an article $\nu$, its ``Fashion DNA'' vector $f_\nu$, is obtained as the activation in a low-dimensional ``bottleneck'' layer near the top of the network.  At its base, the network receives the catalog information as its input: RGB image data is first processed with a pretrained residual neural network \cite{resnet} whose output is concatenated with the categorical and numerical article data and further transformed with a stack of fully connected layers, resulting in Fashion DNA.  As we are ultimately interested in customer preferences, it is sensible to train the model on the sales record:  Disregarding the timestamp information, we arrange the sales information for a large number of frequent customers ($\sim$100k) into a sparse binary purchase matrix $\Pi$ whose elements $\Pi_{\nu k} \in \{ 0,\,1 \}$ indicate whether customer $k$ has bought item $\nu$.  The network is then trained to minimize the average cross-entropy loss per article over these customers.  In effect, the network learns both an optimal representation of the article $f_\nu$ across the customer base, and a logistic regression from Fashion DNA to the sales record for each customer $k$, with weight vectors $s_k$ and bias $\beta_k$ that encode their style preferences and purchase propensity, respectively.  The model architecture is sketched in Figure~\ref{fig:static_net}.

The result is a low-rank logistic factorization of the purchase matrix akin to collaborative filtering \cite{kdd:matrixfactorization,kdd:logisticfactorization}, 
\begin{equation}
\label{static_prob}
\Pi_{\nu k} \, \approx \, p_{\nu k} \,=\, \sigma\left( f_\nu \cdot s_k + \beta_k \right) \;,
\end{equation}
(where $\sigma(\cdot)$ denotes the logistic function), except that the Fashion DNA $f_\nu$ is now clamped to the catalog data via the encoding neural network.  This is a decisive advantage for our setting where we are faced with a continuously changing inventory of goods, as the Fashion DNA for new articles is obtained from their curated data by a simple forward pass through the neural network.  

Ranking the purchase probabilities $p_{\nu k}$ in Eq.~(\ref{static_prob}) naturally induces recommendations \cite{kdd}, a model we use for comparison in Section~\ref{subsec:static}.  We emphasize that the lack of time of sale information enforces static customer styles.  Hence, to invoke dynamically evolving customer tastes, we have to modify the style vectors $s_k$.

\subsection{LSTM network for purchase sequences}
\label{subsec:lstm}
Fashion~DNA provides a compact encoding of all available content information of an item, and largely solves the cold-start problem for new articles entering the store. For these reasons, we use the static model Fashion DNA as article representations in the dynamic model.  We also want to preserve the association between customer-item affinity, and the scalar product of Fashion DNA and customer style, akin to Eq.~(\ref{static_prob}).  Hence, we make our model dynamic by allowing the customer style to change over time $t$.  To distinguish between static and dynamic customer styles, we denote the latter $d_k(t)$.

While we could add time as a dimension to the static model, and attempt to factorize the resulting three-dimensional purchase data tensor (as is done, for example, in \cite{temporal}), we chose to follow a different approach featuring LSTM cells. We also reverse the role of articles and customers:  While our implementation of the static model used batches of articles as input, and learned the response of all customers simultaneously, the input to the LSTM network is customer based. Batches now contain Fashion DNA sequences of the form $( f_{k,1},\ldots, f_{k,N_k} )$, representing the purchase history $\nu_{k,1}, \ldots, \nu_{k,N_k}$ of customer $k$.  When customers buy multiple items at once, the purchase sequence is ambiguous. To prevent the LSTM from interpreting these non-sequential parts as time series, we put purchases with the same time stamp in random order.  Beyond the order sequence, the absolute time of purchases $t_{k,1}, \ldots, t_{k,N_k}$ carries important context information for our problem.  For example, the model may use temporal data to infer the in-store availability of an article, and the season.  We thus additionally supply the time stamp of each purchase to the network.

\begin{figure}
\centering
\includegraphics[width=0.97\columnwidth]{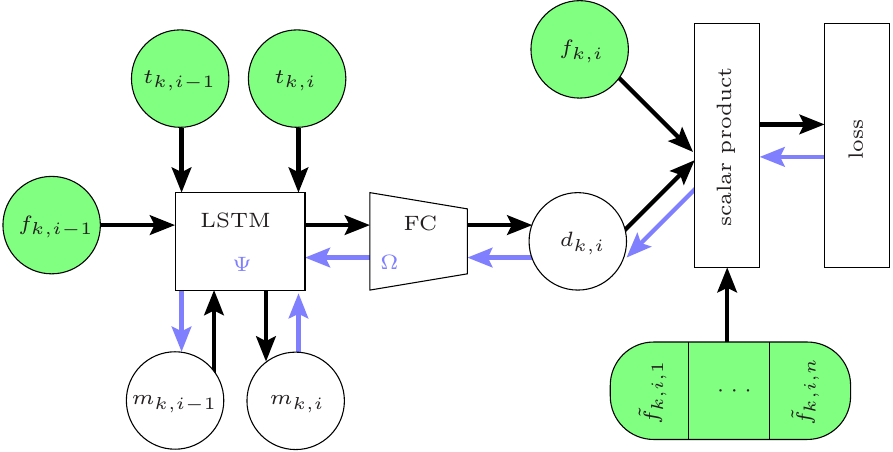}
\caption{Training the dynamical model. The shown time-instance of the LSTM communicates with earlier instances via the memory cells $m_{k,i-1}$ and $m_{k,i}$. They trigger backpropagation through time (blue arrows).}
\label{fig:lstm_net}
\end{figure}

A single pass of the LSTM network processing customer purchase histories is illustrated in Figure~\ref{fig:lstm_net}. For a fixed customer $k$ and purchase number $i$, the LSTM takes as input the concatenation of the time stamp $t_{k,i-1}$ and Fashion~DNA $f_{k,i-1}$ of the previous purchase, and the time stamp $t_{k,i}$ of the current purchase. In addition, the LSTM accesses the content of its own memory, $m_{k,i-1}$, which stores information on the purchase history of customer $k$ it has seen so far. The output of the LSTM is projected by a fully connected layer which results in the current customer style $d_{k,i}$. Note that the first purchase of the sequence ($i=1$) is treated specially: Since there is no previous purchase, we flush $f_{k,0}$, $t_{k,0}$, and $m_{k,0}$ with zero entries. Consequently, the customer style $d_{k,1}$ just depends on the time stamp $t_{k,1}$ and favors the most popular items at that time.

\subsection{Training scheme}
\label{subsec:scheme}
For recommendation, we aim to predict customer style vectors $d_{k,i}$ that maximize the affinity $f_{k,i}{\cdot} d_{k,i}$ to the next-bought article, while minimizing the affinity to all other items in store at that time.  Because it is expensive to compute the customer affinities for every article, we only pick a small sample of ``negative'' examples among the articles not bought.  We denote their corresponding Fashion~DNA vectors by $\tilde{f}_{k,i,1}, \ldots, \tilde{f}_{k,i,n}$. The number of negative examples $n>0$ is a hyperparameter of the model.

We tested three choices of loss functions for training the network, sigmoid cross-entropy loss $\mathcal{L}_\sigma$ (as in the static model), softmax loss $\mathcal{L}_\mathrm{smax}$, and sigmoid-rank loss $\mathcal{L}_\mathrm{rank}$~\cite{rank_loss}, and varied the number $n$ of negative examples. The loss functions are given by:
\begin{equation}\label{eq:loss}
\setlength{\arraycolsep}{0.5ex}
\begin{array}{rcl}
\mathcal{L}_\sigma & = & 
  -\,\mathrm{log}\;\sigma\left(f_{k,i} \cdot d_{k,i}\right) - 
  \sum\limits_{j=1}^n \mathrm{log}\;\sigma\left(- \tilde{f}_{k,i,j} \cdot d_{k,i}\right)\;, \\
\mathcal{L}_\mathrm{smax} & = & 
  -\,\mathrm{log}\left( \frac{\mathrm{exp}\left(f_{k,i} \cdot d_{k,i}\right)}
    {\mathrm{exp}\left(f_{k,i} \cdot d_{k,i}\right)+\sum\limits_{j=1}^n \mathrm{exp}\left(\tilde{f}_{k,i,j} \cdot d_{k,i}\right)} \right)\;, \\
\mathcal{L}_\mathrm{rank} & = & \frac{1}{n}
  \sum\limits_{j=1}^n \sigma\left(\tilde{f}_{k,i,j} \cdot d_{k,i} - f_{k,i} \cdot d_{k,i}\right)\;.
\end{array}
\end{equation}
Only $\mathcal{L}_\mathrm{smax}$ permits a probabilistic interpretation of the dynamical model (when $n$ reaches the number of all available articles).

The minimization landscape for $\mathcal{L}_\sigma$ and $\mathcal{L}_\mathrm{smax}$ depends on the number of negative examples, as their contribution to the loss increases with $n$. Our experiments show that recommendation quality improves when we use more negative examples. Yet, no significant additional benefit is observed when $n$ exceeds $50$.  In contrast, $n$ has no effect on the minimization landscape for the sigmoid-rank loss. Still, for larger $n$ fewer training epochs are needed to adjust the network parameters. We find that $n=20$ is a good tradeoff between faster convergence of the weights, and the computational costs caused by using more negative examples.

A subtle yet important aspect of the recommendation problem is that we try to predict items in the {\em next order} of the customer, rather than inferring articles {\em within a single order}.  As items that are bought together tend to be related (consider, e.g., a swimwear top and bottom), an LSTM network trained on full purchase sequences quickly focuses on multiple orders and overfits.  To circumvent the problem, we let only the first article in the purchase sequence contribute to the loss when a multiple order is encountered.  (Because purchases with the same time stamp are always shuffled before feeding, the LSTM receives a variety of article sequences during training.)

\subsection{Inference and ranking}
\label{subsec:inference}
For each customer $k$, we now define an ``intent-of-purchase'' $\mathrm{ip}_{\nu, k}(t)$ for all articles $\nu$ in store at time $t$, akin to Eq.~(\ref{static_prob}):
\begin{equation}
\label{intent_purchase}
\mathrm{ip}_{\nu, k}(t) \,=\, f_\nu \cdot d_k(t) \;.
\end{equation}
Here, $d_k(t)$ is the dynamic style vector emitted by the LSTM network after feeding {\em all sales} to customer $k$ that occurred before the time $t$ (with randomly assigned sequence for items purchased together); for the final sale, we replace the time stamp of the next purchase by the evaluation time $t$.  We note that $\mathrm{ip}_{\nu, k}(t)$, unlike $p_{\nu k}$~(\ref{static_prob}), cannot be interpreted as a likelihood of sale.  

\section{Comparison of models}
\label{sec:comparison}
To evaluate our dynamic customer model, we assembled sales data from the online fashion store for an eight day period immediately following training, July 1--8, 2015.  We identified customers with orders during this test interval, representing ${\sim}10^5$ individual sales, among $\sim$190k items that were available for purchase in at least one size, for at least one day in this period.  For comparison, we score also the static recommendation model (Section~\ref{subsec:fdna}), and a simple empirical baseline that disregards customer specifics.

\subsection{Empirical baseline}
\label{subsec:baseline}
Fashion articles in the Zalando catalog vary greatly in popularity, with few articles representing most of the sales.  This skewed distribution enables a simple, non-personalized baseline recommender that projects the recent popularity of items into the future.  In detail, we accumulated article sales for the week immediately preceding the evaluation interval (June 23--30, 2015), and defined a popularity score for each article by their sales count if they were still available after July 1.  For those articles (re-)entering inventory during the evaluation period, we assigned the average number of sales among all articles as a preliminary score.  The empirical baseline model then ranks the articles by descending popularity score.

\subsection{Static Fashion DNA model}
\label{subsec:static}
The Fashion DNA network (Section~\ref{subsec:fdna}) provides the basis for a more sophisticated, personalized recommender system, based on the customer static style vectors $s_k$ and the predicted probability of purchase $p_{\nu k}$ (\ref{static_prob}), as detailed in Ref.~\cite{kdd}.  Indeed, $p_{\nu k}$ proves to be an unbiased estimate for the probability of purchase {\em over the lifetime of customer and article}.  These assumptions are not met here, because the evaluation interval is outside the training period, and lasts only eight days.  Still, we may assume that the inner products $f_\nu \cdot s_k$ underlying Eq.~(\ref{static_prob}) are a measure of the affinity of an individual customer $k$ to the in-store items $\{\nu\}(t)$ during the time of evaluation, and sort them by decreasing value to create a static article ranking.

\subsection{Dynamic recommender system}
\label{subsec:dynamic}
For the dynamic customer model, we rank the in-store articles for each customer $k$ according to their intent-of-purchase $\mathrm{ip}_{\nu, k}(t_k)$, see \eqref{intent_purchase}, evaluated at the time of first sale $t_k$ during the evaluation period.  We experimented with the three loss models detailed in Section~\ref{subsec:scheme}, and found comparable results for the sigmoid cross-entropy loss $\mathcal{L}_\sigma$ and sigmoid-rank loss $\mathcal{L}_\mathrm{rank}$, while the softmax loss $\mathcal{L}_\mathrm{smax}$ performed significantly worse.
The following results are based on a pretrained $128$-float Fashion~DNA and an LSTM implementation with $256$ cells, sigmoid-rank loss and $n=20$ negative examples. %
Note that $1 - \mathcal{L}_\mathrm{rank}$ provides a smooth approximation for the area under the ROC curve \cite{rank_loss2}, used for model evaluation below.

\subsection{Results}
\label{subsec:results}
\begin{figure}
\centering
\includegraphics[width=0.97\columnwidth]{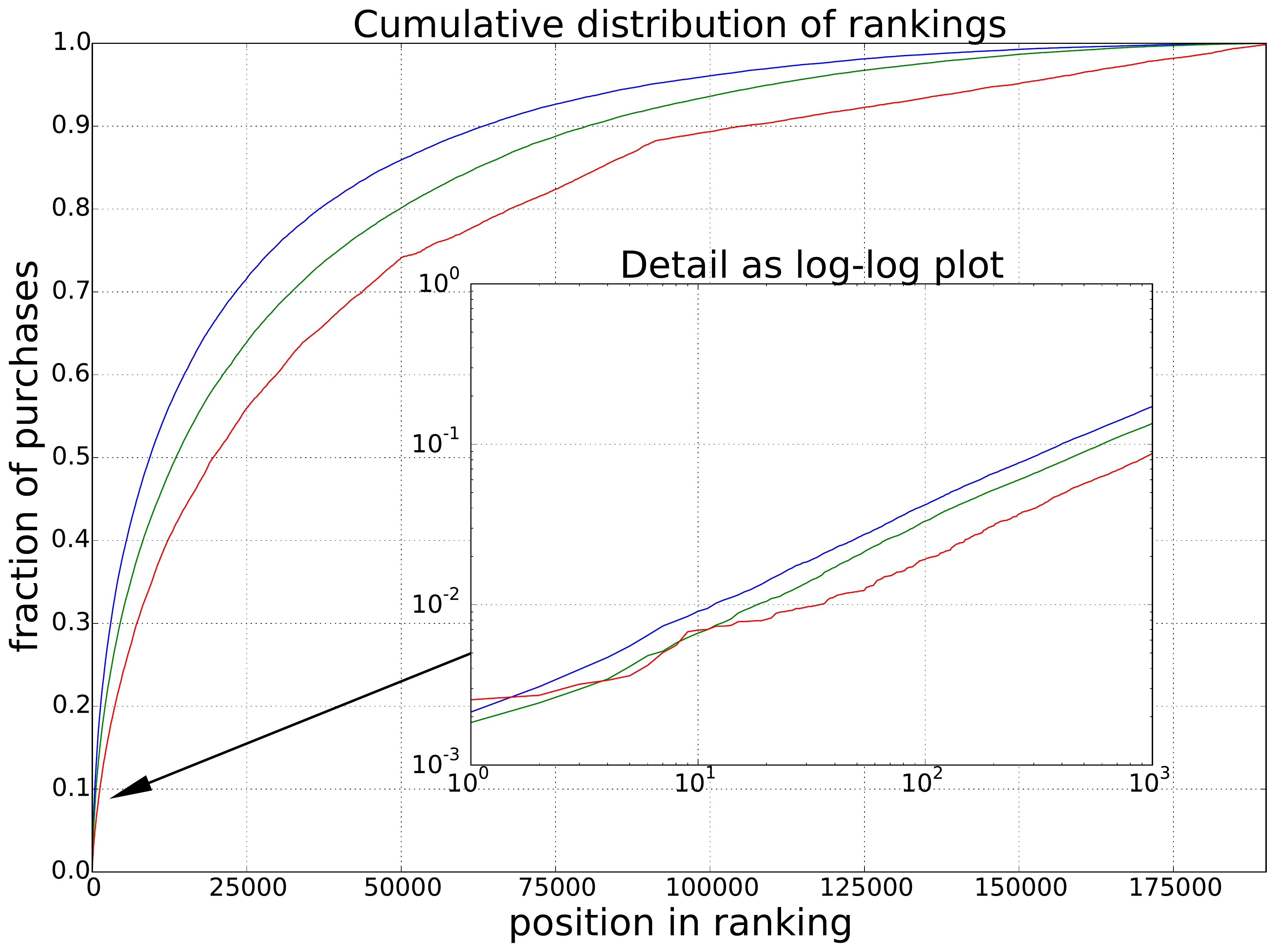}
\caption{ROC curves for the dynamic (blue), static (green), and empirical baseline (red) recommender schemes.}  
\label{fig:comparison}
\end{figure}

To compare model performance, we compile recommendation rankings of the $z \approx 190\mathrm{k}$ items in store for each customer (for the baseline, the ranking is shared among customers), and identify the positions $r_{\nu k}$ of the articles $\{\nu\}(k)$ purchased by customer $k$ during evaluation.  We then determine the cumulative distribution of ranks:
\begin{equation}
\label{cumulative_rank}
R_j \,=\, \sum\nolimits_k \sum\nolimits_{\nu \in \{\nu\}(k)} \mathop{H}\left( j - r_{\nu k} \right) \;.
\end{equation}
$\mathop{H}(\cdot)$ denotes the Heaviside function.  The normalized cumulative rank $R_j / R_z$ interpolates among customers and serves as a collective receiver operating characteristic (ROC) of the recommender schemes (Figure~\ref{fig:comparison}). The inset displays a double-logarithmic detail of the origin region, representing high-quality recommendations.

Table~\ref{tab:comparison} lists the area under the curves (AUC) as a global performance measure, together with quantiles of the distributions $R_j$.  We find that our dynamic model outperforms the static model throughout, and both models are superior to the baseline popularity model, except for the leading $\sim$10 recommendations, representing less than $0.5\%$ of the purchases (inset in Figure~\ref{fig:comparison}).
The table also lists the number of model parameters.  Weights are shared among customers for the LSTM network, but not the static model, resulting in reduction of complexity by orders of magnitude.

More than $3\%$ of the purchased articles from the test interval have not been sold before and, hence, were completely ignored during training. For those \textit{new} articles, the cold start problem applies and the AUC of the baseline, static, and dynamic model decreases to $64.4\%$, $83.3\%$, and $87.7\%$, respectively. In comparison to the numbers displayed in Table~\ref{tab:comparison}, the baseline shows a drastic performance drop as would also be expected from any other recommender system solely based on collaborative filtering. 
Static and dynamic model, however, circumvent this problem thanks to Fashion~DNA.
\renewcommand{\arraystretch}{1.1}
\begin{table}
\centering
\caption{Model comparison. AUC and required number of recommendations to cover 10\% (50\%, 90\%) of purchases.}\label{tab:comparison}
\begin{tabular}{|r|r|r|r|r|r|}
\hline
\textbf{model} & \textbf{AUC} & \textbf{10\%} & \textbf{50\%} & \textbf{90\%} & \textbf{\#params} \\
\hline
 baseline & 80.2\% & 1,200 & 19,500 & 105,000 & - \\
\hline
static & 85.2\% & 600 & 13,500 & 80,000 & $\sim 10^8$\\
\hline
dynamic & \textbf{88.5\%} & \textbf{400} & \textbf{9,300} & \textbf{63,000} & $< 10^6$\\
\hline
\end{tabular}
\end{table}
\section{Outlook}
\label{sec:outlook}
We find that a personalized recommendation model, based on a recurrent network, outperforms a static customer model in the fashion context.  By encoding temporal awareness into the LSTM memory of the network, the dynamic model can infer the seasonality of items, and also record when certain articles are trending---a distinct advantage over the static model, which is limited to learning only long-term customer style preferences.

An important element currently missing in the recommendation model is short-term customer intent.  In the fashion setting, goods for sale belong to varied classes (clothes, shoes, accessories, etc.), and shoppers, irrespective of their style profile, often have a particular category in mind during a session.  These implicit interests strongly influence item preference, but due to their transient nature, are hard to infer from the purchase record.  Complementary data sources like search queries, or the sequence of items viewed online, will pick up the relevant signals instead.  Models that successfully integrate long-term style evolution and short-term customer intent promise to greatly enhance recommendation quality and relevance, and we plan to investigate them in future studies.


\begin{thebibliography}{10}

\bibitem{kdd}
C.~Bracher,  S.~Heinz,  and  R.~Vollgraf.
\newblock Fashion {D}{N}{A}:  Merging  content  and  sales  data  for  recommendation and article mapping.
\newblock In {\em Workshop Machine learning meets fashion}, KDD, 2016.

\bibitem{Devooght}
R.~Devooght and H.~Bersini.
\newblock Long and Short-Term Recommendations with Recurrent Neural Networks.
\newblock {\em Proceedings of the 25th Conference on User Modeling, Adaptation and Personalization} (2017), pp. 13--21.

\bibitem{deeplearning}
I.~Goodfellow, Y.~Bengio, and A.~Courville.
\newblock {\em Deep learning}.
\newblock MIT Press (Cambridge, Mass., USA), 2017. 

\bibitem{resnet}
K.~He, X.~Zhang, S.~Ren, and J.~Sun.
\newblock Deep residual learning for image recognition.
\newblock {\em CoRR} abs/1512.03385 (2015).

\bibitem{rank_loss2}
A.~Herschtal and B.~Raskutti.
\newblock Optimising area under the {R}{O}{C} curve using gradient descent.
\newblock {\em ICML: Conference Proceedings} (2004), pp. 49--.

\bibitem{lstm}
S.~Hochreiter and J.~Schmidhuber.
\newblock Long short-term memory.
\newblock {\em Neural Comput.}~{\bf 9} (1997), p.~1735--1780.

\bibitem{kdd:logisticfactorization}
C.~Johnson.
\newblock Logistic matrix factorization for implicit feedback data.
\newblock In {\em NIPS Workshop on Distributed Matrix Computations}, 2014.

\bibitem{Ko}
Y.--J.~Ko, L.~Maystre, and M.~Grossglauser.
\newblock Collaborative recurrent neural networks for dynamic recommender systems.
\newblock {\em JMLR: Workshop and Conference Proceedings}~{\bf 63} (2016), p.~366--381.

\bibitem{kdd:matrixfactorization}
Y.~Koren, R.~Bell, and C.~Volinsky.
\newblock Matrix factorization techniques for recommender systems.
\newblock {\em IEEE Computer}~{\bf 42} (2009), p.~30--37.

\bibitem{Wang}
H. Wang, X. Shi, and D. Yeung.
\newblock Collaborative recurrent autoencoder: recommend while learning to fill in the blanks.
\newblock {\em Advances in Neural Information Processing Systems}~{\bf 29} (2016), pp. 415--423.

\bibitem{temporal}
L.~Xiong, X.~Chen, T.--K.~Huang, J.~Schneider and J.~G.~Carbonell.
\newblock Temporal collaborative filtering with {B}ayesian probabilistic tensor factorization.
\newblock {\em Proceedings of the 2010 SIAM International Conference on Data Mining} (2010), pp. 211--222.

\bibitem{rank_loss}
L.~Yan, R.~Dodier, M.~C.~Mozer, and R.~Wolniewicz.
\newblock Optimizing classifier performance via approximation to the {W}ilcoxon--{M}ann--{W}itney statistic.
\newblock {\em ICML: Conference Proceedings} (2003), pp. 848--855.

\end{thebibliography}
\end{document}